\documentclass[manuscript]{raa}            % referee version: for submission
\usepackage{graphicx,times}             %for PS/EPS graphics inclusion, new
\usepackage{natbib}
\usepackage{amssymb,amsmath}
\bibpunct{(}{)}{;}{a}{}{,}

\usepackage[a4paper=true,dvipdfm=true,pagebackref=true]{hyperref}
\hypersetup{colorlinks = true, linkcolor = green, anchorcolor = red, citecolor = blue, filecolor = red, pagecolor = red, urlcolor = red}

\begin{document}

   \title{Some interesting topics provoked by the solar filament research in the past decade
%\footnotetext{$*$ Supported by the National Natural Science Foundation of China.}
}
%   \subtitle{I. Place Your Subtitle Here}

   \volnopage{Vol.0 (20xx) No.0, 000--000}      %%preserved for Editor. DOn't remove!
   \setcounter{page}{1}          %%starting page, preserved for Editor. DOn't remove!

   \author{Peng-Fei Chen \inst{1,2}
      \and Ao-Ao Xu   \inst{3}
      \and Ming-De Ding   \inst{1,2}
   }

   \institute{School of Astronomy and Space Science, Nanjing University,
Nanjing 210023, China; {\it chenpf@nju.edu.cn}
        \and
        Key Laboratory of Modern Astronomy \& Astrophysics, Nanjing University, China\\
	\and
	State Key Laboratory of Lunar and Planetary Sciences, Macau University of Science and Technology, Macau, China\\
\vs\no
   {\small Received~~2020 Sep 1; accepted~~2020~~Oct 8}}

\abstract{Solar filaments are an intriguing phenomenon, like cool clouds
suspended in the hot corona. Similar structures exist in the intergalactic
medium as well. Despite being a long-studied topic, solar filaments have
continually attracted intensive attention because of their link to the coronal
heating, coronal seismology, solar flares, and coronal mass ejections (CMEs). In
this review paper, by combing through the solar filament-related work done in the
past decade, we discuss several controversial topics, such as the fine
structures, dynamics, magnetic configurations, and helicity of filaments. With
high-resolution and high-sensitivity observations, combined with numerical
simulations, it is expected that resolving these disputes will definitely lead
to a huge leap in understanding the physics related to solar filaments, and
even shed light on galactic filaments.
}

   \authorrunning{P.-F. Chen, A. A. Xu \& M.-D. Ding}   %author_head in even pages
   \titlerunning{Interesting topics about solar filaments}% title_head in odd pages

   \maketitle
%% The author head (on even pages) and the title head (on odd pages) will be
%% automatically extracted from \author{} and \title{}. Whenever the title is too long,
%% you will be asked to supply a shorter one by inserting either \authorrunning{} or
%% \titlerunning{} before \maketitle. Anyway, you can specify your own heads.
%%
%%
%% Note: In the following text body of your manuscript, please note several differences from
%%       other major journals:
%% (1) \subsection{Please Capitalize the First Letter of Each Notional Word in Subsection Title}
%% (2) Please Capitalize the First Letter of Each Notional Word in all tables' captions

%
%________________________________________________ sections below
%
\section{Introduction}           %% first-level sections will be auto-capitalized
\label{sect:intro}

As the closest star to humans, the Sun not only provides the providential heat
and light for the lives on Earth, but also exhibits a variety of intriguing
activity, initially to the naked eyes and later through telescopes. Sunspots own
the longest written record, which was made by Chinese in 165 BC \citep{witt87} or
even 800 BC. Solar filaments might come off second earliest, whose discovery was
dated back to 1239 \citep{tand74}.

Solar filaments are elongated dark features against the bright solar disk in
various wavelengths such as H$\alpha$, H$\beta$, \ion{He}{i} 10830 \AA, \ion{Ca}
8542 \AA, \ion{Ca}{ii} H \& K, \ion{Na}{D1, D2 \& D3}, as well as \ion{He}{i}
304 \AA. These are typical spectral lines formed at the chromospheric
temperatures, matching the thermal properties in filaments. Solar filaments are
also identifiable as dark features in extreme ultraviolet (EUV) images, which is
mainly attributed to volume blocking \citep{heinzel03}. When solar filaments move
to the solar limb following the solar rotation, they are revealed to be
suspended above the solar limb. In this case, they are called prominences. Solar
filaments and prominences are generally used exchangeably in the literature.

Talking about solar prominences, we have to mention that historically any plasma structure
standing out of the solar limb was called a prominence \citep{dejag59, zirin88},
including surges, spicules and postflare loops. In the modern semantic environments, these
dynamic phenomena are not classified as prominences. Another similar phenomenon
is the so-called arch filament systems \citep{suyn18}, which look very similar to
solar filaments in H$\alpha$ images, consisting of many threads. However, arch
filament systems are short-lived dynamic structures associated with emerging
magnetic flux.
As a result, they are very different from solar filaments in the Dopplergrams:
Their central part displays blueshifts and the footpoints of their threads display
redshifts. Moreover, one key difference between solar filaments and arch filament
systems is that the threads in the former are generally weakly skewed from the
magnetic neutral line, whereas the threads in arch filament systems are
quasi-perpendicular to the underlying magnetic neutral line.

Another issue worth clarifying is whether solar filaments should be called a
chromospheric structure or a coronal structure, both of which were used in the
literature. The temperature of filament plasma is in the range of 6,000--8,000 K,
which is the typical temperature of the solar chromosphere. This is why solar
filaments are most discernable in the H$\alpha$ line, which is widely used to
observe the chromosphere. However, we are strongly against the statement
describing solar filaments as chromospheric structures. We stress here that solar
filaments are cold structures suspended in the hot corona and thus are coronal
structures. A filament might be rooted at the solar chromosphere, it
might also hang totally well above the solar chromosphere.

Filaments are a fascinating phenomenon in the solar atmosphere not only for their
stunning appearance, but also for the physics involved in the whole lifelong
evolution: (1) Their formation is related to thermal instability, which is
fighting all the time with coronal heating. We tend to believe that the
secret about how
the corona is heated is partly hidden in the process how the coronal plasma cools
down to form filaments. In particular, the thermal discontinuities brought by
filaments provide a favorite environment for magnetic energy deposit
\citep{low15}. (2) Their oscillations and dynamics can be utilized to
diagnose the coronal magnetic field that cannot be measured precisely to date
\citep{arre18}. (3) Their eruptions are intimately related to solar flares and
coronal mass ejections \citep[CMEs,][]{chenpf11}, the two major eruptions in the
solar atmosphere that might pose disastrous disturbances to the space environment
near the Earth. This is why it was stated that solar filaments, once erupting,
are not only the core of CMEs, but also the core of CME research
\citep{chenpf14}. From the longer timescale point of view, solar filaments trace
the magnetic neutral lines associated with active regions and decayed active
regions, their latitude evolution displays a butterfly diagram similar to
sunspots \citep{haoq15}, and they show the same hemispheric asymmetry as sunspots
\citep{likj10, kongdf15}. Because of the importance of solar filaments, many
monographs and review papers were devoted to this topic \citep{paren14, vial15,
gibson18}. In this review, rather than covering every aspect of the filament
research, we will focus mainly on some new or controversial topics that were
provoked in the past decade and are in strong need of clarification.

This paper is organized as follows. After compiling what we have known about
solar filament in Section \ref{sect:known}, we discuss in detail those debating
issues or newly proposed ideas in the solar filament research in Section
\ref{sect:topics}, and some less touched but worthwhile topics are mentioned in
Section \ref{sect:future}.

\section{What we have already known}
\label{sect:known}

\subsection{Filament Formation}

While the cosmic background radiation cools down monotonically after the big bang,
the baryonic matter in the universe struggles between superhot ($10^6$--$10^7$ K)
states and supercold ($10$--$10^3$ K) owing to the competitive interplay between
various kinds of heating and runaway radiative cooling, making mixing structures
with different sizes. On large scales like tens of kpc, one example of the mixture
of cold and hot plasmas is the cold galactic filaments embedded in the hot
galactic corona. On smaller scales like tens of Mm, the typical example is the
solar filaments, which correspond to $\sim$7000 K plasma embedded in the $10^6$ K
solar corona.

Early in 1950s, \citet{parker53} proposed that the cold filaments are formed due
to thermal instability where thermal conduction is the restoring process. Two
factors contribute to such a thermal instability. First, the radiative cooling
is proportional to $n_e^2$, whereas the coronal heating is proportional to $n_e$
or insensitive to $n_e$ depending on the heating mechanisms, where $n_e$ is the
electron density in the corona. Therefore, a perturbation with a negative
$\Delta T$ and positive $\Delta n$ (keeping the gas pressure unchanged) would
lead to further cooling. Second, the radiative loss function increases with the
decreasing temperature $T$ in the range of $T>2\times 10^5$ K. It means
that a perturbation with a negative $\Delta T$ in the $10^6$ K corona enhances
radiative cooling, and a runaway instability happens.

For two reasons it is believed that the filament material does not come from the
quiet corona itself. First, the element abundance of many filaments is similar to that
of the low solar atmosphere and different from that of the corona
\citep{spicer98, songhq17}. This means that the plasma in the photosphere and
chromosphere should somehow fill in the filaments that are high up in the
corona. Second, the plasma density inside the filament is $\sim$100
times higher than that of the ambient corona, and the typical length of a
filament thread is $\sim$10 Mm \citep{liny05}. This means that the magnetic flux
tube should be larger than $10^3$ Mm, which is almost 10 times the typical
magnetic field length in filament channels. It implies that there is no
sufficient mass in the coronal portion of the in situ magnetic flux tubes to
feed a long filament thread. It is noted in passing
that it was often claimed that the mass of 10 quiescent filaments is roughly the
total mass of the entire corona \citep{malherbe89}, which might not be precise.
The typical mass of a filament is $\sim$$10^{14}$--$2\times 10^{15}$
\citep{paren14}, whereas the mass of the corona from the bottom to an altitude
of 0.2$R_\odot$ amounts to $\sim 6\times 10^{17}$ g.

The straightforward solution is that the chromospheric cool plasma is heated
locally to evaporate into the corona, where the hot and dense plasma cools down
owing to thermal instability. Such a mechanism is sometimes called
chromospheric evaporation--coronal condensation model or
evaporation--condensation model for short \citep{anti00}. It was further found
that if the ratio of apex to footpoint heating rates is less than $\sim$0.1 and
asymmetries in heating and/or cross-sectional area are less than $\sim$3, no
equilibrium can be reached in the whole magnetic loop, which is called thermal
non-equilibrium. The corona either keeps evolving in a limit cycle with changing
temperatures or condenses into a filament near magnetic dips \citep{klim19}. In
the latter case, it should be the thermal instability that leads the coronal
plasma to a lower temperature rather than a higher temperature. The condensation
process was verified by the observations of the Atomospheric Imaging Assembly
(AIA) aboard the Solar Dynamics Observatory \citep[SDO, ][]{berger12,
liuw12}. Besides the chromospheric evaporation, which triggers the thermal
instability via enhancing plasma density, it is interesting to note that
adiabatic expansion of the coronal loops might do the same job by decreasing the
temperature of the coronal loop \citep{lee95}.

The second mechanism of filament formation is the direct injection of the
chromospheric plasma into the corona \citep{wang99, wangjc18, wangjc19}.
Different from the evaporation--condensation model or the thermal non-equilibrium
model, where filaments appear in the corona from nothing in H$\alpha$
observations, the injection mechanism is manifested as direct injection of
H$\alpha$ surges across the filament channel. In this scenario, the filament
threads might be quasi-perpendicular to the underlying magnetic neutral line,
rather than quasi-parallel with the magnetic neutral line, as seen from Figure
7 of \citet{zou16}.

The third formation mechanism is the so-called levitation model, where
chromopheric plasma is lifted along with emerging flux \citep{rust94}, which
was occasionally observed \citep{xuz12}. Magnetohydrodynamic (MHD) simulations
indicated that chromospheric magnetic reconnection might facilitate the
levitation of the heavy filaments \citep{zhaoxz17}. If the emerging flux is a
flux rope, we expect to see an arch filament system (which corresponds to the
top part of the flux rope), followed by a filament. The threads in the arch
filament system and the filament should be oppositely skewed from the magnetic
neutral line.

It should be kept in mind that the Sun is always more complicated than we think.
Filaments might be formed in a way quite different from the above-mentioned
models. For example, it was recently proposed that magnetic reconnection in the
low corona can also trigger thermal instability, leading to filament formation
\citep{kaneko17}. Such a process was confirmed by SDO/AIA observations
\citep{lilp19}. In this case, the mass source is the corona itself, and no
chromospheric evaporation is needed (unless reconnection-accelerated electrons
bombard the chromosphere to generate chromospheric evaporation). It is expected
that the element abundance in this type of filaments should be the same as that
of the corona. Moreover, it is noticed that there are ubiquitous chromospheric
fibrils, among which the fibrils near sunspots are longer than those in quiet
regions \citep{jing19}.
Filaments can also be formed due to the interactions between neighboring fibrils
or between a fibril and a superpenumbral filament, where magnetic reconnection
was proposed to be the key \citep{biy12, xuezk16, yanxl16, yangb16b, yangb16a}.

\subsection{Filament Oscillations}

Any object in equilibrium is subject to external perturbations, and may
inevitably deviate from the equilibrium position. If the restoring force is
strong enough, the object can come back, leading to oscillations. With various
types of energy dissipation mechanisms such as viscosity and radiation, the
kinetic energy, which can be as large as $10^{26}$ erg \citep{shenyd14a}, would
finally damp out. Therefore, oscillations can provide vital information on the
triggering process, restoring force, and damping mechanisms. Moreover, filament
oscillations without significant decay might be a precursor for filament
eruptions and CMEs \citep{chenpf08, lit12, zhengrs17}. As a result, there has
been booming research on filament/prominence oscillations, and prominence
seismology has been progressing rapidly \citep{arre18}.

Based on the nature of the restoring force, filament oscillations can be divided
into longitudinal oscillations and transverse oscillations. In the former case,
the filament plasmas oscillate along the magnetic field lines, hence the Lorentz
force $\mathbf{J}\times \mathbf{B}$ does not affect the motion, and the restoring
force includes the field-aligned component of gravity and the gas pressure
difference between the two ends of the filament thread \citep{luna12, zhangqm12}.
It was found that the gravity component is the dominant one unless the magnetic
dip is too shallow. In the case of transverse oscillations, the plasma and the
frozen-in magnetic field move perpendicular to the magnetic field, where Lorentz
force takes effect \citep{shenyd14b, zhouyh18}. As a result, the period of
longitudinal oscillations is generally of the order of 1 hr, whereas that of
transverse oscillations is around $\sim$20 min. Caution is warranted that the
period for each mode has a wide range.

It is not as straightforward as we thought to identify the oscillation mode
\citep{pant15, pant16, chenjl17}. Whether it is longitudinal or transverse
depends on whether the plasma motion is along or perpendicular
to the magnetic field. There was misunderstanding in the literature
claiming that longitudinal oscillations is along the filament axis or attributing
any lateral displacement to transverse oscillations. Noticing that the magnetic
field lines follow the threads, which deviate from the filament axis (or spine)
by $10^\circ$--$30^\circ$ \citep{hana17}, even longitudinal oscillations would
present displacement perpendicular to the filament spine.

Depending on the attack angle relative to the filament threads, a large-scale
coronal shock wave might cause longitudinal oscillations in one filament and
transverse oscillations in the other \citep{shenyd14b}, or even no response from
a low-lying filament \citep{liur13}. The co-existence of longitudinal and
transverse oscillations in one filament recently attracted much attention
\citep{pant16, wangb16, zhangqm17a, mazu20}. Whether the two modes are excited
separately or there is mode conversion is definitely another interesting topic
\citep{liak20}.

Once an oscillation mode is determined, its period can be used to diagnose the
magnetic field strength and configuration in the corona. For example, in terms of
longitudinal oscillations, the period was applied to derive the curvature radius
of the local magnetic dip \citep{luna12, zhangqm12, zhouyh18}. The curvature
evolution can also be deciphered, e.g., when a filament is activated to rise
slowly, its longitudinal oscillation period was observed to increase, implying
that the magnetic dip became flatter and the flux rope became less twisted
\citep{biy14}. It was also observed that the oscillation period increases in some
threads but decreases in other threads of the same filament, implying correlated
magnetic rearrangement \citep{zhangqm17b}. In terms of transverse oscillations,
the period was applied to estimate the magnetic field strength around the
filaments \citep{zhouyh18}. For example, \citet{zhangqm18} applied the seismology
to an oscillating prominence and found that the magnetic field in the cavity is
less than 10 G.

As an important ingredient of prominence seismology, the decay time of filament
oscillations also discloses vital information. For longitudinal oscillations, the
simulation results of \citet{zhangqm12, zhangqm20} indicated that radiative
cooling and thermal conduction are not sufficient to explain the decay, and extra
factors should be taken into account. Extra damping mechanisms include mass
drainage \citep{zhangqm13}, mass accretion \citep{ruder16}, wave leakage
\citep{zhangly19}, and increase of the background coronal temperature
\citep{ruder16}. The longitudinal oscillations can be amplified by additional
perturbations \citep{zhangqm20} or the decrease of the background coronal
temperature \citep{ruder16}. For transverse oscillations, \citet{adro20}
proposed that resonance absorption is the main mechanism. Wave leakage might
play a role as well.

\subsection{Filament Eruptions}

Filaments end their lives by either thermal disappearance or eruptions. Their
thermal disappearance might be due to enhanced coronal heating or mass drainage
without further mass replenishment. Their eruptions can be successful or failed
ones. For successful eruptions, the whole picture can be explained by the
standard flare/CME model, where magnetic reconnection plays a crucial role
\citep{chenpf11}. Although the model holds true in principle, it is basically
a schematic sketch, with many details waiting to be supplemented. In the past
decade, most advances were concentrated on the new characteristics brought by
3-dimensional (3D) magnetic reconnection compared to its 2-dimensional (2D)
counterpart
\citep{meizx17} and some new features brought by the complex background magnetic
field. For example, non-uniform or reconnection-favored background magnetic field
(including multiple spine-fan configurations) might lead to rich dynamics
of the ejecta that cannot be accounted for in the original standard model, such as
the deflection, splitting, disintegration, mass transfer, and even rotation
\citep{biy13, wangym16, yangjy15, lilp16, chenhc18, liur18, lihd19, weih20,
yanxl20}. A hot channel is also formed due to magnetic reconnection, which
maps the flux rope with the erupting cool filament at the bottom
\citep{chengx14b}. We tend to link this hot channel to the fuzzy component
observed in the CME core \citep{songhq19}.

More attention was paid to the triggering mechanisms \citep{zhangqh20}, which can
be divided into ideal MHD type and reconnection type. Sometimes the two types of
processes might work together \citep{songhq15}. In the past decade, plenty of
efforts were made on ideal MHD mechanisms \citep{biy15, meizx18}, such as kink
instability and torus instability. Contradictory results were obtained on whether
the threshold of the torus instability is similar \citep{xingc18} or different
\citep{zoup19} between active-region filaments and quiescent filaments. One
uncertain issue in these works is whether slow magnetic reconnection is already
going on during the claimed ideal MHD triggering process. A possible signature
of such slow reconnection is the appearance of the EUV hot channel before the
impulsive phase of the associated flare in many events. Noticing that torus
instability was sometimes considered to be the driving mechanism for the full
eruption rather than a triggering mechanism \citep[see][for a review]{schm15},
\citet{chenpf19} doubted this by a logic test: Suppose the decay index of the
background magnetic field exceeds the threshold of the torus instability
everywhere from the low corona all the way to the interplanetary space, can a 3D
flux rope erupt to form a CME purely due to torus instability? If it could, this
process would violate the Aly-Sturrock constraint.

Several possible reasons for erupting filaments to fail were explored. They can
be hindered by overlying arcades \citep{chenhd13}, gravity \citep{fili20},
filament rotation \citep{zhouzj19}, or another filament lying above
\citep{jiangyc14}. There are also events where a part of the filament erupts
successfully and the other part fails \citep{zhangqm15}.

\section{Some new or controversial topics}
\label{sect:topics}

\subsection{Fine Structures}

   \begin{figure}
   \centering
   \includegraphics[width=\textwidth, angle=0]{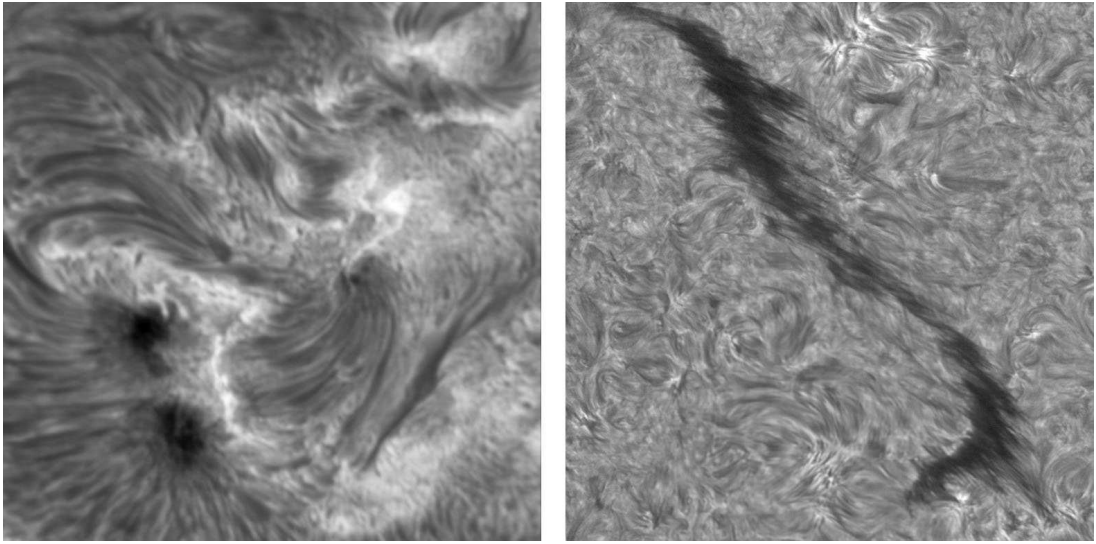}
   \caption{Solar filaments observed by the New Vacuum Solar Telescope in
	   H$\alpha$. Left panel: several active region filaments around two
	   sunspots; Right panel: a quiescent filament.}
   \label{fine}
   \end{figure}

Figure \ref{fine} shows several active region filaments (left panel) and a quiescent
filament (right panel), both of which were observed by the Chinese New Vacuum
Solar Telescope \citep[NVST,][]{liuz14}. Fine structures can be seen clearly. It
is generally claimed that a filament is composed of a spine and a few barbs. We
have to stress here that the spine is not a real structure, at least not for a
quiescent filament. As indicated by the right panel of Figure \ref{fine},
filament threads are actually the building blocks of a filament. The illusion of
a spine is simply due to the fact that all the threads are situated above the
magnetic neutral line. Our statement may hold true even for active-region
filaments, though their threads are nearly parallel with the magnetic neutral
line.

\begin{figure}
   \centering
   \includegraphics[width=\textwidth, angle=0]{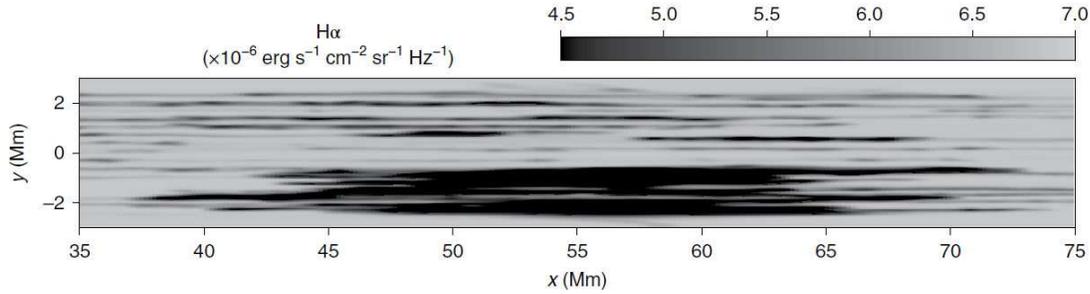}
   \caption{Synthesized H$\alpha$ image of a filament in MHD simulations, showing
	   a filament is composed of many thin threads. Taken from
   \citet{zhouyh20}.  }
   \label{zhouyh}
   \end{figure}

Although it is well known that threads are the building blocks of filaments
\citep{liny05}, one question that had not been asked is why filaments exist in
the form of a collection of threads, rather than being clumpy. To answer this
question, \citet{zhouyh20} performed 2D MHD simulations with a high
spatial resolution, by which they proposed that turbulent heating in the solar
surface drives random chromospheric evaporation, leading to sporadic
coronal condensation in some magnetic dips. Once a thread is born, its gas
pressure decreases as demonstrated by \citet{xiac11}, which results in the
shrinkage of the local flux tube and the expansion of the neighboring flux
tubes (hence low density). It implies that once a thread is formed, its
neighboring flux tubes
become unfavored space for coronal condensation. They also found that the
filling factor of threads in a filament channel is proportional to the
strength of the chromospheric heating, which can explain why some filaments
look seamless, in particular those filaments in active regions, where the
heating is much stronger than in quiet regions. As shown in Figure
\ref{zhouyh}, the threads have a length from several to 30 Mm, and an average
width of 100 km. They occupy 10--15\% space of the filament channel. All these
are consistent with observations \citep{liny05}.

\begin{figure}
   \centering
   \includegraphics[width=6cm, angle=0]{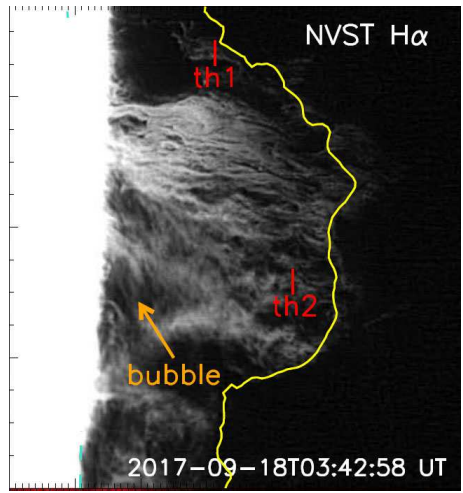}
   \caption{A prominence above the solar limb observed by NVST in H$\alpha$,
	  showing vertical threads, plumes, and bubbles. Taken from
   \citet{lid18}.  }
   \label{lid}
   \end{figure}

Prominences, especially quiescent prominences, show more complicated fine
structures, with many vertical threads as well as dark plumes emerging from the
low-lying bubbles, as seen in Figure \ref{lid} \citep{berger11, lid18}. It is
still unclear how these vertical
threads are formed. Are their magnetic field lines oriented vertically or mainly
horizontal but turbulent? \citet{xiac16b} performed 3D MHD simulations, and
proposed that magnetic Rayleigh-Taylor instability turns an initially horizontal
magnetic field into turbulent, forming vertical threads. A parade of vertical
threads forms apprently horizontal threads. In this model, the horizontal threads
are apparent structures. \citet{schm14} and \citet{ruangp18} held a different
viewpoint, claiming that the magnetic field is mainly horizontal, and the vertical
threads in quiescent prominences are apparent structures due to piling up of a
series of small dips. With coronal magnetic extrapolations, \citet{suyn15} found
that the vertical threads in a polar crown prominence is supported by
horizontal dips in a flux rope, implying that the field lines crossing the
prominence are horizontal. There is a similar debate on whether the filament
threads on the solar disk are field-aligned \citep{zhouyh20} or field-misaligned
\citep{claes20}.

\subsection{Chirality and Helicity}

Similar to many other natural phenomena like the typhoon on the Earth, solar
filaments also possess chirality. From the magnetic perspective, imagine
that we stand on the positive polarity side of a filament channel, if the axial
magnetic field in the filament is toward left, the filament is called sinistral;
if the axial magnetic field in the filament is toward right, the filament is
called dextral \citep{martin92}. The corresponding current helicity
$H_c=\mathbf{J}\cdot \mathbf{B}$ is positive/negative in sinistral/dextral
filaments. From the H$\alpha$ or EUV image perspective, filament barbs are
either left-bearing (like the highway exits in Japan) or right-bearing (like the
highway exits in China). \citet{martin98} proposed a correlation between these
two types of chirality, i.e., left-bearing/right-bearing barbs correspond to
sinistral/dextral filaments. In another word, filaments with
left-bearing/right-bearing barbs have positive/negative helicity. This
correspondence provides a handy approach to identify the sign of helicity simply
from H$\alpha$ or EUV images. As a result, this approach has been widely applied
to investigate the hemispheric rule of helicity.

However, \citet{guoy10} found both left- and right-bearing barbs in one
filament with negative helicity. Their coronal magnetic extrapolation indicated
that the filament segment with the left-bearing barb is supported by a sheared
arcade, whereas the filament segment with the right-bearing barb is supported by
a flux rope. Along this line of thought, \citet{chenpf14} put forward that
Martin's rule is correct only for the filaments supported by flux ropes, and the
correspondence between the barb bearing and the helicity sign is opposite for
the filaments supported by sheared arcades. They noticed that once a filament
erupts, some materials drain down to the solar surface, producing twin
brightenings on the two sides of the magnetic neutral line. The twin drainage
sites also show chirality, i.e., they are either left skewed or right skewed
relative to the magnetic neutral line. \citet{chenpf14} proposed that left-skewed
twin drainage sites correspond to negative helicity and right-skewed drainage
sites correspond to positive helicity. A similar scheme was already proposed by
\citet{wangym09}, though with a distinct explanation. Applying this approach to
571 erupting filaments observed by SDO/AIA during 2010 May--2015 December,
\citet{ouy17} found that 91.6\% of these filaments follow the hemispheric rule
of helicity sign, i.e., negative/positive in the northern/southern hemisphere.
They also investigated the cyclic evolution of such a hemispheric rule, and
found that the hemispheric preference of quiescent and intermediate filaments
keeps significant in the whole solar cycle, but that of active-region filaments
can disappear near solar maximum or minimum. The coronal helicity is
contributed by both magnetic flux emergence and surface motions. The
significant hemispheric preference of the quiescent and intermediate filaments
tends to support the helicity condensation model \citep{anti13, kniz17}, where
helicity injected from the solar surface due to small-scale rotational flows
inversely cascades from small scales to large scales. The rotational flows are
supposed to be driven by the Coriolis force acted on the granular or
supergranular flows.

\subsection{Magnetic Configuration}

   \begin{figure}
   \centering
   \includegraphics[width=12cm, angle=0]{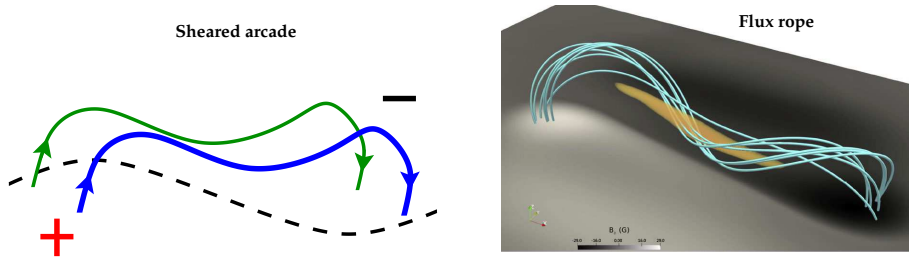}
   \caption{Two types of magnetic configurations for solar filaments. Left
	   panel: A sheared arcade with normal polarity dips; Right panel:
	   A flux rope with inverse-polarity dips. The right panel is taken
	   from \citet{zhouyh18}.  }
   \label{2config}
   \end{figure}

The magnetic configuration of solar filaments is an extremely important research
topic since filaments are the source regions of many flare/CME events. Before
going to the details, we first comment on the question whether it is necessary
for a filament to be supported by magnetic dips, which were originally proposed
to balance the gravity last century. With numerical simulations, \citet{karp01}
demonstrated that even in a simple magnetic loop without any dips, chromospheric
evaporation--coronal condensation can also happen, leading to repeated thread
forming and draining. Besides, for the filaments owing to the injection
mechanism \citep{wang99}, cold chromospheric material is injected into the coronal
loop from one end and drains down to the other end of the magnetic field line.
Hence, magnetic dips are not necessary either. Unfortunately, the coronal magnetic
field is still an important parameter that we cannot measure directly with
sufficiently high precision. However, one important clue from observations is
that if we can observe longitudinal oscillations (including the small-amplitude
counterstreamings) in the filament threads, we can judge that magnetic dips exist
in these filaments. It is reminded that we cannot simply exclude magnetic dips
when longitudinal oscillations are not observed. Although no survey has been
taken to determine the percentage of the filaments with magnetic
dips, our impression based on the above clue is that a large number of filaments
are supported by magnetic dips. Therefore, we focus on these filaments in the
rest of this subsection.

   \begin{figure}
   \centering
   \includegraphics[width=12cm, angle=0]{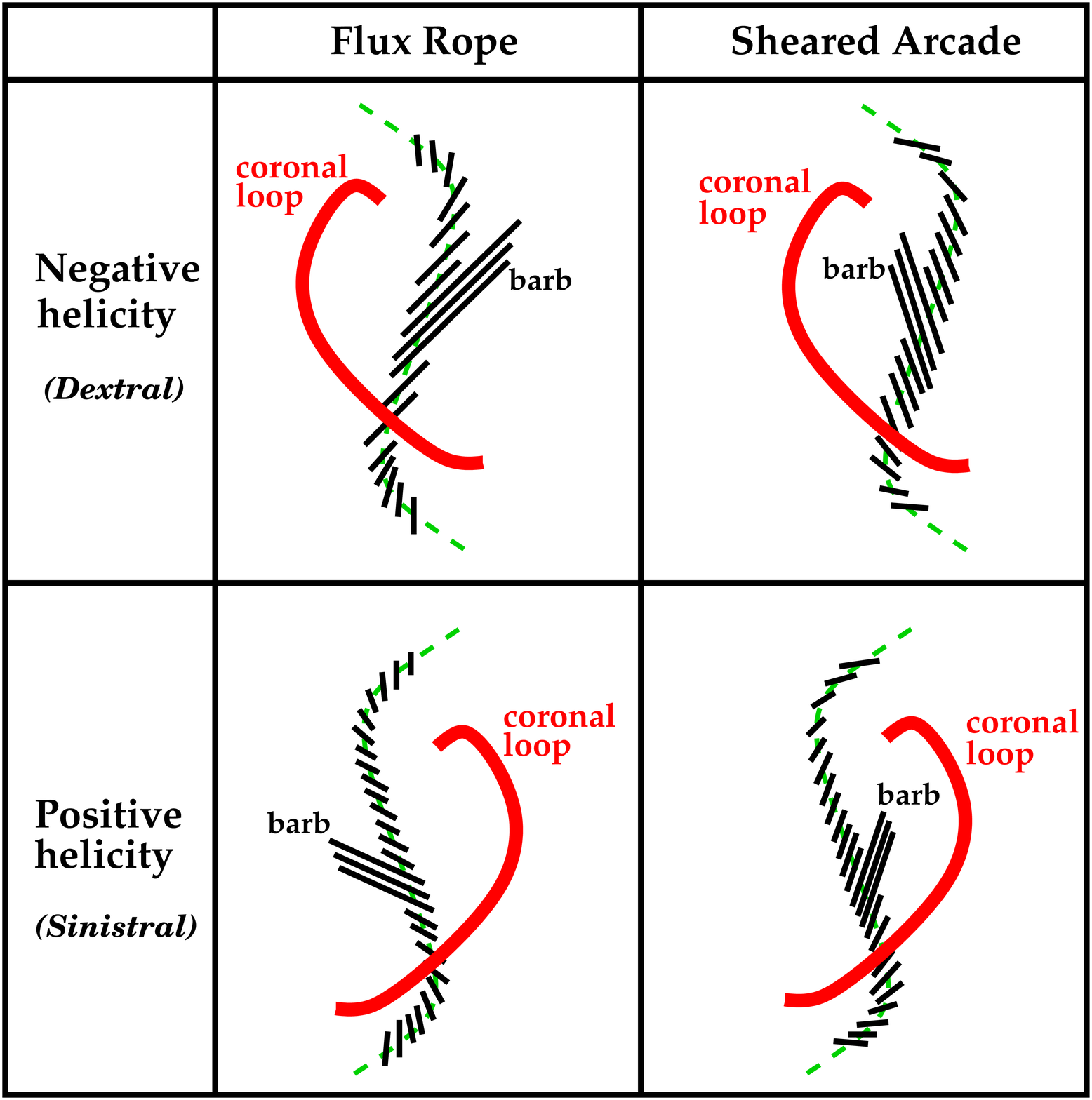}
   \caption{Schematic sketch demonstrating the correspondence between the helicity
	   and the chirality of the filament barbs (longer black lines), threads
	   (shorter black lines), and the overlying coronal loops (red lines) in
	   two magnetic configurations, i.e., flux ropes and sheared arcades. The
	   green dashed lines mark the magnetic neutral lines of filament
	   channels.}
   \label{chen14}
   \end{figure}

Suppose a filament channel with positive magnetic polarity on the left side and
negative polarity on the right side, there are two types of magnetic dips above
the magnetic neutral line in the corona: normal-polarity dips with magnetic field
pointing to the right and inverse-polarity dips with magnetic field pointing to
the left. The typical corresponding magnetic configurations are a sheared arcade and a flux
rope, which are displayed in Figure \ref{2config}. As mentioned in \citet[][and
references therein]{gibson18}, there is also a normal-polarity flux rope model.
However, this model was sketched in 2D. When mapped to 3D space where the flux
rope is rooted to the solar surface, there arises a problem of mixed chirality.

There were sporadic measurements of magnetic field inside prominences, which are
not routinely available nowadays. However, \citet{chenpf14} proposed an indirect
way to distinguish these two types of magnetic configurations based on EUV or H$\alpha$ 
images only: If the helicity of a filament channel is positive/negative and the
filament barbs or threads are left/right bearing, then the filament is supported
by a flux rope; If the helicity of a filament channel is positive/negative and the
filament barbs or threads are right/left bearing, then the filament is supported
by a sheared arcade. This method is illustrated in Figure \ref{chen14}, according
to which the famous scheme proposed by \citet{martin98} holds true only for the
filaments supported by flux ropes. Applying this method to 571 filaments observed
by SDO/AIA, \citet{ouy15, ouy17} found that 89\% of the filaments
are supported by flux ropes and 11\% are by sheared arcades. Such a ratio is
similar to that obtained from coronal magnetic extrapolations \citep{duan19}.

The widely used method to decipher the filament magnetic configuration is still
the magnetic extrapolation based on photospheric vector magnetograms
\citep{suyn15, guoy17, guoy19, mack20, zhuxs20}. Whereas sheared arcades were
paid less attention, most papers were devoted to the events with flux ropes
\citep[see][for reviews]{chengx17, liur20}. For these papers, while many
extrapolations showed that the flux ropes before eruption are generally weakly
twisted \citep{jiangcw14, lisw17}, several authors claimed that highly twisted
flux ropes can nicely match the filament geometry or spectro-polarimetric
observations \citep{guoy19, mack20}. The corresponding twist is larger than 3
turns, which is well above the threshold of kink instability \citep{hood81}. If
highly-twisted flux ropes can be stable, one possible reason might be that the
gravity hinders the filaments from erupting \citep{fanyh20}, which significantly
increases the threshold of kink instability. Highly-twisted flux ropes imply that
there are multiple dips, hence multiple threads, along one flux tube, which
brings new topics, for example, the thread longitudinal oscillations are not
independent any more, and there is thread-thread
interaction, which significantly changes the damping time of the oscillations,
sometimes leading to decayless oscillations \citep{zhouyh17}. The signatures
of such thread-thread interaction was found in observations \citep{zhangqm17b}.

With high-resolution and high-sensitivity observations, the magnetic
configurations of some filaments can be mapped by tracing the plasma motions
when filaments are activated. In this way, the twist of the coronal magnetic
field can be estimated, which was found to range from $1\pi$ to $6\pi$ in
different events \citep{yanxl14, yangsh14, houyj16, chenhd19, shenyd19,
xuhq20}. The lower limit might correspond to a sheared arcade, while others
to a flux rope.

It is noted that not all magnetic configurations can be exclusively classified
as a sheared arcade or a flux rope. A filament might be supported by a flux
rope in a segment and by a sheared arcade in the other \citep{guoy10}.
Besides, the existence of double-decker filaments requires a combination of
the two elementary configurations, i.e., sheared arcade plus sheared arcade,
flux rope plus flux rope, and sheared arcade plus flux rope. The method
proposed by \citet{chenpf14}, see Figure \ref{chen14}, can be used to discriminate these possibilities.
The double-decker filaments may erupt in different ways depending on the
background magnetic field \citep{kliem14}: Maybe the upper branch erupts,
leaving the lower branch almost
intact \citep{chengx14a, zhengrs19}, or the lower branch erupts first, which then
pushes the upper branch to erupt after coalescence \citep{zhucm15}. Interestingly,
there are events with magnetic configurations similar to the double-decker
filaments, but no cool material exists in the upper branch of the magnetic
structure. In this case, as the upper branch erupts, an erupting hot channel is
visible, but the lower filament remains intact \citep{liut18}.

\subsection{How Filament Barbs Are Formed}

As indicated by Figure \ref{fine}, filament barbs are indeed an eminent ingredient
of a filament, which veer away from the apparent spine. Generally, active region
filaments have less barbs (some even have no barbs), whereas quiescent filaments
have more barbs. According to the statistical research of \citet{haoq15}, about
75\% filaments have less than five barbs, and a small portion of filaments have
more than ten barbs. \citet{lilp13} studied a polar crown filament, which has 69
barbs. According to their results, the formation of filament barbs is associated with 
three types of plasma motions, and the disappearance of barbs is also
accompanied by three types of plasma motions that are influenced by the
changing photospheric magnetic field.

   \begin{figure}
   \centering
   \includegraphics[width=12cm, angle=0]{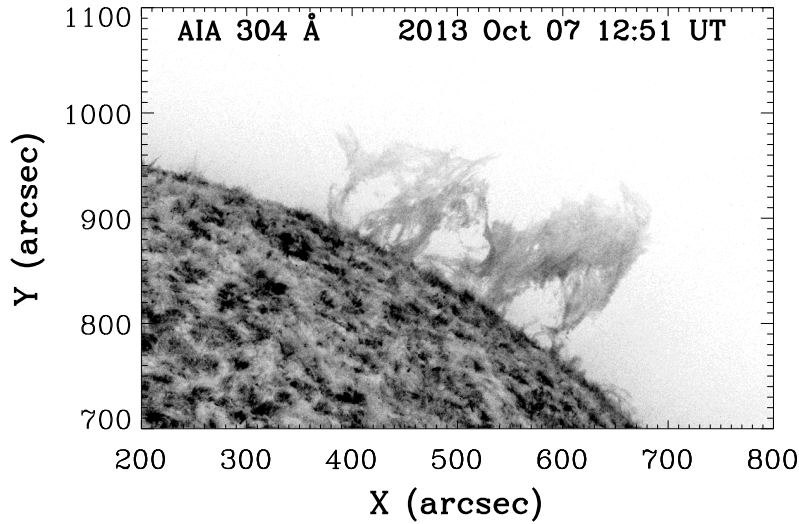}
   \caption{A solar prominence observed by SDO/AIA at 304 \AA\ on 2013 October
	   7, showing several feet attached to the solar surface. This figure
	   is a negative image.}
   \label{foot}
   \end{figure}

When a filament lies above the solar limb as a prominence, it often has several
feet, as shown in Figure \ref{foot}. It is generally taken for granted that a
filament barb corresponds to the projection of a lateral foot, which extends down
to the solar surface. With a linear force-free magnetic field extrapolation,
\citet{aulan98} proposed that the intrusion of a parasitic polarity into a bipolar
filament channel would produce a series of new magnetic dips, which extends from
the main flux rope to the solar surface. If these new magnetic dips are filled
with cold plasma, they match the observed filament barbs very well when seen from
above. Such a lateral dip assemble model can explain many filament barbs
\citep{chae05}.

Interestingly, recent observations indicated that there might exist another type
of barbs, which are due to longitudinal oscillations of some threads in the
filament, hence were called ``dynamic barbs" \citep{awas19}. With quadrature
observations from SDO and STEREO-B, \citet{ouy20} confirmed that the dynamic
barbs are due to longitudinal oscillations, and the barb does not extend down to
the solar surface. Hence, they proposed that a filament barb does not always
correspond to a prominence foot. Similarly, when some threads drain down along
the field line, a transient barb is also visible \citep{xiac14}.

Besides, if the magnetic dips are not identical, with some dips much longer than
others, their threads would be much longer than other threads as well since
the length of a thread is proportional to the length of the magnetic dip
\citep{zhouyh14}. In this case, a barb would also be present without the help of
parasitic magnetic polarity in the filament channel. Since this type of barbs
are simply due to local longer threads, it is expected that such filament barbs
do not correspond to prominence feet as well.

\subsection{The Nature of Counterstreamings}\label{counter}

Similar to the quiet Sun, which is never really quiet, any filament is never
static even when the whole appearance looks invariant with time. Flows and
small-scale oscillations are ubiquitous. It is estimated that filament material
drains down to the solar surface with a rate of $\sim$$10^{15}$ g day$^{-1}$
\citep{liuw12}, implying that the plasma of a filament is recycled completely
in the timescale of $\sim$1 day. MHD simulations verified such mass cycling
\citep{xiac16a}.

Spectroscopic observations indicated that redshifted and blueshifted motions are
pervasive in the filament threads \citep{schm91}, which were later called
counterstreaming flows \citep{zirk98}. The classical model attributed them to
filament thread longitudinal oscillations \citep{liny05}. Indeed, different
magnetic dips have different curvature radii, leading to different oscillation
periods according to the pendulum model. Moreover, perturbations imposed on
individual thread are often random. These two factors result in the longitudinal
oscillations out of phase, i.e., some are forward and some are backward, forming
counterstreamings.

On the other hand, \citet{chenpf14} proposed that the counterstreaming flows have
another component, i.e., unidirectional flows with opposite directions. This
scenario was confirmed by various observations \citep{yanxl15, lit16, zou16,
wanghm18}. Therefore, it is expected some counterstreamings are due to filament
thread longitudinal oscillations, some are due to unidirectional flows, and some
are due to their combination as demonstrated in numerical simulations
\citep{zhouyh20} and observations \citep{dier18, wanghm18, pane20}. It is also
conceivable that for the filaments formed via the injection mechanism
\citep{wang99}, their counterstreamings tend to be mainly due to unidirectional
flows. Furthermore, \citet{zhouyh20} proposed that there exist hot
counterstreaming flows in the interthread corona, which is purely due to
unidirectional flows.

The counterstreamings mentioned above are horizontal flows along magnetic field
lines, therefore they are mostly observed in filaments on the solar disk as
proper motions or in prominences above the solar limb as Doppler-shifted patterns.
With high-resolution observations, vertical threads in the hedgerow
prominences also present complicated upward and downward motions
\citep{berger11}, which were called vertical counterstreamings by
\citet{shenyd15}. This type of upward and downward flows are also visible in
filament barbs, and their formation was suggested to be a mixture of simultaneous
flows, waves, and magnetic field motions \citep{lilp13, kucera18}. This is
definitely a topic deserving further research.

\subsection{The Nature of Solar Tornados}

Generally speaking, filaments on the solar disk show much simpler morphology
compared to prominences above the solar limb, even when the spatial resolution
of the observations is not high. Such an overlooked characteristic might
actually have a profound implication: threads, the building blocks of filaments,
are generally situated on horizontal magnetic flux tubes. Therefore, when
viewed from above, we can see the whole thread, and all the quasi-parallel
threads lining up along the magnetic neutral line show simple morphologies.
When viewed from the side as in the case of prominences, the thread assemble
might make up any pattern due to the projection effects, especially when the
threads are roughly along the line of sight. As a result, even in 1930s,
\citet{pett32} classified solar prominences into 5 classes, one of which is
tornado-like.

With the high-resolution, SDO/AIA observations revealed many tornado-like
structures not only at the bottom of a coronal cavity \citep{lix12} but also in
filament feet/barbs \citep{suy12}.
\citet{lix12} interpreted the vortex-like motions as a combination of mass flows
and density waves propagating along the helical magnetic fields, whereas
\citet{suy12} interpreted the tornado feet of the filaments as rotating magnetic
structures driven by the underlying vortex flows on the solar surface.
\citet{suy12} argued that solar tornados play an important role in supplying mass
and magnetic twists into the
filaments. In this expalanation, the filament feet are similar to rotating
macrospicules, which were also called solar tornados \citep{pike98}. Later,
\citet{suy14} investigated the dynamics of the immediately ambient coronal plasma
around a tornado with the Hinode/EIS spectral data. With opposite velocities
of $\sim$5 km s$^{-1}$ persisting for more than 3 hr on the two sides of the
tornado, they concluded that the blue- and red-shifted motions across the tornado
are consistent with their rotation model. One query this model is confronted
with is whether the persistent twisting motion would transfer too much twist into
the filament body so that the filament becomes kink unstable.

Several authors proposed
alternative explanations. \citet{pana14} suggested that the apparent rotation
motions in the solar tornados are illusion due to counterstreamings of the threads
inside the filament feet. In this case, each thread in the filament foot
experiences longitudinal oscillations around the local magnetic dip, and all the
longitudinal oscillations out of phase would give an illusion that the filament
foot is rotating. Spectroscopic observations can definitely provide additional
information in discriminating apparent motions from mass motions, hence can shed
light on the nature of the solar tornados. \citet{schm17} observed a tornado with
the MSDP spectrograph, and the derived H$\alpha$ Doppler maps show a pattern with
alternatively blueshifted and redshifted areas of 5--10\arcsec\ wide. More
importantly, the blue- and red-shifts change sign with a quasi-periodicity of
40--60 min, supporting the idea that the threads inside the tornado are
oscillating along the basically horizontal magnetic dips, which are mainly
oriented along the line of sight. The longitudinal
oscillations out of phase in different magnetic dips lead to the
counterstreamings and the illusion of rotation. It is noted that the oscillation
period is typical for filament longitudinal oscillations, where gravity serves as
the restoring force \citep{luna12, zhangqm12}.

   \begin{figure}
   \centering
   \includegraphics[width=12cm, angle=0]{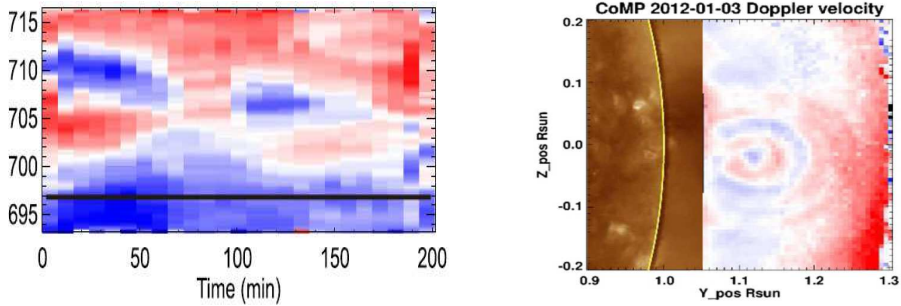}
   \caption{Left panel: Time-distance diagram of the Doppler velocity along a
	   slice across a tornado studied by \citet{yangzh18}. Right panel:
	   A composite image of SDO/AIA 193 \AA\ image (left portion) and
	   the Dopplergram of a coronal cavity (right portion) taken and adapted
	   from \citet{bak13}. }
   \label{yangzh}
   \end{figure}

While the longitudinal oscillations of the threads inside the filament feet are
a sounding explanation in avoiding the over-accumulation of magnetic twist in the
filaments, spectral observations, however, did not always show alternating blue-
and red-shifts quasi-periodically. For example, \citet{yangzh18} used IRIS data
to analyze the profiles of the \ion{Mg}{ii} k 2796 \AA\ and \ion{Si}{iv} 1393
\AA\ lines, which correspond to the cool plasmas with temperatures of $10^4$
K--$10^5$ K. As shown in the left panel of Figure \ref{yangzh}, although some
small patches did show alternating blue- and red-shifts, they found coherent and
stable redshifts and blueshifts across the tornado axis for more than 2.5 hr,
which is several times longer than the typical period of filament barb
longitudinal oscillations \citep{lilp13}. Therefore, they tend to favor these
tornadoes as flowing cool plasmas along a relatively stable helical magnetic
structure. In our view, while their persistent blue- and red-shifts across the
tornado axis definitely cannot be explained by the thread longitudinal
oscillation model, it might still be consistent with the counterstreaming model.
As we discussed in Section \ref{counter}, there are two types of
counterstreamings, i.e., longitudinal oscillations of the cold threads out of
phase and spatially alternating unidirectional flows. The latter, which is due
to siphon flows or jets, can last for a long time, as proposed by
\citet{chenpf14} and observationally confirmed by \citet{lit16}. Similar
persistent forward and backward flows on the two edges of a filament were
confirmed in 3D MHD simulations \citep{xiac16a, xiac17} and observations
\citep{zou17}. The formation of a tornado associated with coronal sub-jets found
by \citet{chenhd17} might also be indirect evidence of the spatially alternating
unidirectional flows as one formation mechanism of solar tornados. Moreover, the
right panel of Figure \ref{yangzh} displays a spiral pattern in the Dopplergram
revealed by \citet{bak13} and \citet{chenyj18}. We tend to attribute it to
spatially alternating unidirectional flows along a flux rope.

How solar tornados disappear is another topic worth investigating. \citet{lilp13}
found that the disappearance of filament feet (here we presume that all filament
feet are tornados, though rotating motions are not clearly visible in some feet
possibly due to their small sizes) are generally accompanied by the disappearance
of the parasitic magnetic polarity nearby. This is understandable since the
small magnetic dips supporting the foot threads vanish in the magnetic
rearrangement following the disappearance of the parasitic polarity.
Alternatively, \citet{chenhd17} proposed that the disappearance of tornados is 
due to self-reconnection of the tangled magnetic fields, which were twisted by
the photospheric vortex.

It is noted in passing that the solar tornados discussed above are quasi-static
structures in dormant
filaments. Once a filament erupts, continuing magnetic reconnection would increase
the magnetic twist of the supporting magnetic flux rope, no matter whether it
pre-exists or is formed during reconnection \citep{ouy15}. During the eruption,
some filament materials move up and some drain down along the twisted flux rope,
manifested as erupting tornados, which have been revealed by both observations
\citep{wangws17} and MHD numerical simulations \citep{jiangcw18}.

\subsection{Is Magnetic Field Deformable due to Filament Gravity}

Although it has been shown that some filaments, especially quiescent prominences,
are subject to Kelvin-Helmholtz instability \citep{lid18} or Rayleigh-Taylor
instability \citep{keppens14, hillier18}, where magnetic field deforms due to
plasma motion or gravity, it was not rare to be claimed in the literature that
the filament gravity cannot deform magnetic field lines when the plasma $\beta$
in the filaments is small, say 0.01--0.1. \citet{zhouyh18} pointed out that this
is misunderstanding, since the plasma $\beta$ represents the strength of gas
pressure compared to magnetic pressure, and has nothing to do with gravity. In
order to characterize gravity, they proposed another dimensionless parameter,
plasma $\delta$, which is the ratio of gravity to magnetic pressure, i.e.,
$\delta={{\rho gL}\over {B^2/2\mu_0}}=11.5{n \over {10^{11}~\rm{cm}^{-3}}}
{L \over {100~\rm{Mm}}} ({B \over {10~\rm{G}}})^{-2}$, where $n$ is the number
density of hydrogen in the prominence, $L$ is the length of the prominence
thread, and $B$ is the magnetic field. If $\delta$ is much smaller than unity,
the magnetic field is not deformable because of the filament gravity. However,
if $\delta$ is comparable to or larger than unity, the field lines can be
deformed by the weight of filaments.

In the 2D MHD simulations of filament oscillations performed by \citet{zhangly19},
the plasma $\delta$ is $\sim$0.8. It was revealed that the magnetic field
geometry changes with time as the filament oscillates along the field line
\citep[see also][]{kra16}. It is such deformation that generates transverse
oscillations of the ambient magnetic field, leading to wave leakage. The
deformation of magnetic field due to filament gravity was also verified by
observations, as indicated by Figure 5 in \citet{lihb18}.

\section{Prospects of other topics}
\label{sect:future}

Solar filaments/prominences, an object chased also by amateurs, are full of
mysteries. With large-aperture ground-based telescopes being developed and new
wavelengths, in particular Ly$\alpha$, being utilized in space missions like Solar
Orbiter, ASOS and CHASE \citep{fangc19, ganwq19, lic19, lih19, vial19, vour19},
the research on solar filaments is expected to boom in the coming decade. Some
topics worth further exploring are summarized as follows:

\begin{itemize}

\item \underline{Thermal structure:} Spectroscopic observations will be promising
	in diagnosing the thermal structure of filaments, including the
	distributions of temperature, density and velocity, as well as the filling
	factor of the threads \citep{heinzel15, ruangp19}.

\item \underline{Magnetic structure of the filament endpoints:} In the literature
	it is widely believed that the magnetic field at the filament endpoints
	is rooted locally, meaning that the footpoints of a filament and the
	magnetic field line are
	co-spatial on the same magnetic polarity. However, for those filaments
	supported by flux ropes, we tend to think that the local magnetic field
	is a bald patch, hence the magnetic field is mainly horizontal and the
	real footpoint of the magnetic field line is rooted at the opposite
	polarity. As argued by \citet{haoq16}, considering the filament endpoints
	as the root of field lines would lead to mis-identification of the
	filament chirality. Note here that the statement does not hold true for
	the jet-like filaments \citep{wang99}.
	
\item \underline{Filament interactions:} There are many sympathetic eruptions,
	where eruption of one filament triggers the other \citep{jiangyc11,
	yangjy12, lisw17, houyj20}. There are more events where two filament
	interact rather gengtly, leading to rich phenomena \citep{jiangyc13,
	kongdf13, daij18, songzp19}, which can be explored further.

\item \underline{{\it p}-mode waves:} The photospheric {\it p}-mode waves can
	easily leak to the filaments since the field lines are strongly
	oblique. There should exist a relationship between {\it p}-mode waves and
	the filament formation and dynamics \citep{caowd10, lit16}. More
	simulations are encouraged.

\item \underline{3D construction of the height and shape:} Rotation of activated
	filaments can tell the sign of helicity of the embedded magnetic field.
	However, 2D images might cheat our eyes as demonstrated by a famous
	animation picture, where a dancing girl can be recognized to rotate
	clockwise or counterclockwise. In this sense, 3D construction of the filament shapes and motions based on stereoscopic or spectroscopic
	observations are important \citep{lit10, lit11, songhq18, zhouzj20}.

\item \underline{Allowance height:} Observations indicated that most prominences
	have an upper edge below 50 Mm, and the maximum height of the upper edge of quiescent
	prominences is $\sim$500 Mm \citep{wangym10}. For the injection mechanism,
	\citet{zou19a} estimated that the maximum height is about 25 Mm. One
	question can be readily asked: what is the maximum height in the solar
	corona for a prominence to form via the evaporation-condensation
	mechanism? It is reminded that a prominence might be lifted to a higher altitude after birth
	due to quasi-static evolution of the magnetic field. Another question is
	at which height a prominence would become unstable \citep{liuk12}.

\item \underline{Mini-filaments:} It seems that mini-filaments resemble large
	filaments in many ways, including their eruptions as mini-CMEs
	\citep{hongjc11, yangb15, hongjc16} and mini-ICMEs \citep{wangjm19}.
	Future large telescopes can reveal more details in mini-filaments.

\item \underline{Automated data processing:} Over a century-long data archive of
	filament observations has been set up, where H$\alpha$ data from
	worldwide observatories were collected \citep{lingh20}. Owing to the
	huge amount of data, automated recognition or machine-learning methods
	were developed and should be improved further \citep{haoq13, zhugf19},
	in order to statistically study the length, location, orientation, and
	latitude migration \citep{gaopx12}.
%\end{enumerate}
\end{itemize}

\begin{acknowledgements}
Our research was supported by the Chinese foundations (NSFC 11533005,
11961131002, 11733003, and U1731241). X.A.A. was supported by the Science and
Technology Development Fund of Macau (275/2017/A). PFC thanks ISSI-Beijing and
ISSI for supporting team meetings on solar filaments. This paper is dedicated to
Prof. Yunchun Jiang, who trained many students in the filament research.
\end{acknowledgements}

\bibliographystyle{raa}

\bibliography{ms}

\label{lastpage}

\end{document}